\documentclass[aps,prd,preprint,amssymb,nofootinbib,showpacs,preprintnumbers,floatfix]{revtex4-1}

\usepackage{amsmath}
\usepackage{graphicx}
\graphicspath{{./}}

\usepackage{multirow} 

\usepackage[mathcal]{eucal}

\usepackage{epsfig}
\usepackage{dcolumn} 
\usepackage{bm} 
\usepackage{slashed}

\newenvironment{my_itemize}{
\begin{itemize}
  \setlength{\itemsep}{1pt}
  \setlength{\parskip}{0pt}
  \setlength{\parsep}{0pt}}{\end{itemize}
}

\newenvironment{my_enumerate}{
\begin{enumerate}
  \setlength{\itemsep}{1pt}
  \setlength{\parskip}{0pt}
  \setlength{\parsep}{0pt}}{\end{enumerate}
}

\newcommand{\eqnref}[1]{Eq.~(\ref{eqn:#1})}

\newcommand{\secref}[1]{Sec.~\ref{sec:#1}}

\newcommand{\figref}[1]{Fig.~\ref{fig:#1}}

\newcommand{\tableref}[1]{Table~\ref{table:#1}}

\begin{document}

\preprint{UCI-TR-2010-19}

\title{A New Method for Resolving Combinatorial Ambiguities at Hadron
  Colliders}

\author{Arvind Rajaraman}
\email{arajaram@uci.edu}

\author{Felix Yu}
\email{felixy@uci.edu}
\affiliation{Department of Physics and Astronomy,
University of California, Irvine, CA 92697-4575, USA}

\date{September 14, 2010}


\begin{abstract}
We present a new method for resolving combinatorial ambiguities that
arise in multi-particle decay chains at hadron colliders where the
assignment of visible particles to the different decay chains has
ambiguities.  Our method, based on selection cuts favoring high
transverse momentum and low invariant mass pairings, is shown to be
significantly superior to the more traditional hemisphere method for a
large class of decay chains, producing an increase in signal retention
of up to a factor of 2.  This new method can thus greatly reduce the
combinatorial ambiguities of decay chain assignments.

\end{abstract}




\maketitle


\section{Introduction}
\label{sec:introduction}

The turn-on of the Large Hadron Collider ushers in a data-driven era
of particle physics.  Major theoretical frameworks, such as
supersymmetry (SUSY) and universal extra dimensions (UED), will be
tested using present and upcoming collider data.  In many such models,
the lightest new physics particle is protected from decaying into
Standard Model particles because of a discrete  symmetry, e.g.
$R$-parity or Kaluza-Klein parity for SUSY and UED, respectively.  The
resulting lightest supersymmetric particle (LSP) or lightest KK-odd
particle (LKP), if neutral, is usually a viable dark matter candidate.

Measuring the properties of this dark matter candidate is critical for
determining its cosmological behavior and its ultimate suitability as
the dark matter.  Because of the discrete symmetry, these particles
must be produced in pairs at colliders, but since they escape, they
manifest themselves in a collider event as large amounts of missing
transverse momentum.  Since only the total missing transverse momentum
can be measured, the kinematics of these events cannot be fully
reconstructed, and as a result, the dark matter candidate mass cannot
in general be measured in any given event.  It is therefore important
to have methods of measuring this mass, which also serves to set the
overall scale of new physics.

A large number of kinematic analysis techniques have been proposed in
the literature to extract the mass of the dark matter candidate.
Broadly speaking, these can be divided into three categories.  The
first is based on edge measurements of invariant mass distributions,
using the fact that the algebraic expressions for such endpoints are
related to the on-shell masses of the cascade decay
chain~\cite{Baer:1995nq, Hinchliffe:1996iu, Baer:1998sz,
  Hinchliffe:1998ys, Bachacou:1999zb, Hinchliffe:1999zc,
  Allanach:2000kt,Hisano:2002xq, Hisano:2003qu}.  The second class of
analysis methods is known as the polynomial method, which solves the
non-linear kinematic four-momentum conservation equations and thus
determines the masses of the entire decay chain~\cite{Nojiri:2003tu,
  Kawagoe:2004rz, Cheng:2007xv, Cheng:2008mg, Cheng:2009fw}.  The last
broad category of approaches uses new kinematic observables and
functions such as $m_{T2}$, $m_{CT}$, and
$m_{CT2}$~\cite{Lester:1999tx, Barr:2003rg, Lester:2007fq, Cho:2007qv,
  Gripaios:2007is, Barr:2007hy, Cho:2007dh, Nojiri:2008hy,
  Tovey:2008ui, Nojiri:2008vq, Cheng:2008hk, Burns:2008va, Barr:2009jv,
  Cho:2009ve, Kim:2009nq, Barr:2010zj, Blanke:2010cm, Cho:2010vz}.

While these methods work well under idealized conditions, in practice
there can be serious obstacles to implementing them successfully.
Perhaps the most important problem, common to all these approaches, is
the presence of combinatoric backgrounds which occur because the new
particles must be produced in pairs.  For example, consider the
process where two gluinos are produced and decay through a squark to a
neutralino: $\tilde{g} \tilde{g} \rightarrow qqqq \tilde{\chi}_1^0
\tilde{\chi}_1^0$.  We will observe four jets, but we do not know
which were emitted first nor from which gluino each was emitted.
There is thus an ambiguity in reconstructing the event from the
visible particles; we will refer to this as a combinatorial ambiguity.
In general, with pair-produced parents that decay via identical long
cascades, combinatorial ambiguities present a major hurdle in
distinguishing the appropriate assignment of particles to each chain
as well as the unique ordering of these particles, as discussed in a
recent review on mass reconstruction techniques~\cite{Barr:2010zj}.
Wrong assignments can lead to significant suppression of mass peaks
and cause large tails in distributions~\cite{Cheng:2009fw}.  We also
note that understanding these ambiguities is crucial in extracting the
nature of the new physics.

We emphasize that these kinematic reconstruction methods are not
affected equally by wrong combinations.  In fact, the most general
transverse variable based on $m_{T2}$, known as
$m_{TGen}$~\cite{Lester:2007fq}, does not suffer from combinatorial
ambiguities since it explicitly includes a minimization over all
possible decay chain assignments of the observed object momenta.  For
regular $m_{T2}$ studies, the assignment ambiguity, {\textit{i.e.}}
assigning particles to separate decay chains, is relevant, while the
ordering ambiguity, {\textit{i.e.}} the sequence of the particles on
the decay chain, is not important: by construction, the transverse
mass of each decay chain is irrespective of decay chain placement, but
is clearly dependent on decay chain assignments.  The subsystem
$m_{T2}$ reconstruction method~\cite{Nojiri:2008vq, Burns:2008va},
however, is adversely affected by both the assignment ambiguity and
the ordering ambiguity.  Similarly, wrong assignment combinations and
wrong ordering choices degrade the effectiveness of the polynomial
method~\cite{Nojiri:2003tu, Cheng:2008mg, Cheng:2009fw} and can also
worsen results from the kinematic edges approach if the invariant mass
distributions do not encapsulate entire decay chains.

In this paper, we will focus on resolving the combinatorial ambiguity
arising from decay chain assignments, leaving the question of
resolving ordering ambiguities for the future.  Thus, our results
should improve the effectiveness of all of the aforementioned
kinematic reconstruction methods except $m_{TGen}$, which, as noted
above, does not suffer from assignment ambiguities.  Henceforth,
``combinatorial ambiguity'' will refer solely to the decay chain
assignment ambiguity discussed above.

In previous kinematic reconstruction studies, past authors have
designed a variety of methods to address combinatorial ambiguities.
For example, in~\cite{Cheng:2008mg, Cheng:2009fw}, they apply the
polynomial method to pairs of events, and they address combinatorial
ambiguities by favoring, for a given event, the assignment that
maximizes the number of event pairings that give algebraic solutions.
In this way, they hoped to discard wrong combinations of a given event
that would give unphysical solutions when paired with correct
combinations of other events.  In~\cite{Blanke:2010cm}, which used
$m_{T2}$-- and sub-system $m_{T2}$--based methods, they chose the
combinations of jet pairs with smaller invariant masses and smaller
angular separation as well as discarding the largest sub-system
$m_{T2}$ values, arguing that correct jet pairings should be
directionally focused, and high invariant mass jet pairings are more
likely incorrect.  While these representative approaches at reducing
wrong combinations work on an individual analysis basis, we note that
these methods are not interchangeable.  Even though the combinatorial
ambiguity for kinematic reconstruction studies is in general a common
difficulty, many of the specific approaches to resolve such
combinatorial ambiguities used in the literature cannot be generalized
when using more than one kinematic reconstruction technique.

One exception is the hemisphere method, used in~\cite{Cho:2007dh,
  Nojiri:2008hy, Nojiri:2008vq, Kim:2009nq} (we will describe this
method in detail below in~\secref{hemisphere}).  If the parent
particles of each decay chain are strongly boosted, their decay
products will be collimated along the initial momenta. By considering
events with suitably large transverse momenta, one may hope to avoid
combinatorial ambiguities (cf. Sec.~13.4 of~\cite{Ball:2007zza}).
However, this can lead to significant loss in signal statistics.  On
the other hand, the hemisphere method of reducing combinatorial
ambiguities allows for the simultaneous application of multiple
kinematic reconstruction methods, without a specialized approach.

In this paper, we present a new method for resolving such
combinatorial ambiguities in decay chains. Our method is described
in~\secref{ptvm}.  We show that our method can yield highly accurate
assignments of particles to decay chains which have little
contamination from wrong combinations.  We contrast our method with a
parallel study of the efficacy of the more familiar hemisphere method.
Our results show that for cascade decay chains with on-shell mass
resonances, our method improves signal retention up to a factor of 2
over the usual hemisphere method.

We begin by presenting two toy models, which we shall use to compare
the two methods.  In~\secref{models}, we present the model masses and
aspects of the simulation.  In~\secref{hemisphere}, we review the
hemisphere method and give our hemisphere method implementation
details.  We then discuss our new method (which we shall call the
$p_T$ v. $M$ method) in~\secref{ptvm} and describe the specifics of
our procedure in~\secref{ptvmprocedure}.  We compare the two methods
in~\secref{comparison} and conclude in~\secref{conclusion}.

\section{Models}
\label{sec:models}

In this section, we present the toy models used in our analysis.  The
masses of the two models considered are summarized
in~\tableref{model}.  Roughly, Model A mimics a SPS1a-style mass
spectrum, while Model B represents an off-shell squark scenario.

Our aim is to isolate the combinatorial ambiguities arising from pure
signal events when the final state particles are indistinguishable
inside the detector.  We focus solely on gluino pair production with
each gluino decaying to the lightest neutralino and two quarks:
$\tilde{g} \tilde{g} \rightarrow qqqq \tilde{\chi}_1^0
\tilde{\chi}_1^0$.  The intermediate squark is on-shell (Model A) or
off-shell (Model B).  All quarks are treated at parton level, and
effects from initial state radiation or parton showering are not
considered.  These events have three distinct pair-pair combinations.
Isolating which particles arise from the same decay chain is a first
step in any kinematic event analysis.  Since we are tackling the
combinatorial ambiguity that arises even when dealing with only signal
events, we shall not complicate our analysis by adding background
events or multiple production or decay modes.

\begin{table}[tbp]
\centering
\begin{tabular}{|l||c|c|c|c|}
\hline
Model Name       & $\tilde{g}$ (GeV) & $\tilde{q}$ (GeV)
& $\tilde{\chi}_1^0$ (GeV)
& Diquark Invariant Mass edge (GeV) \\
\hline
Model A     &  600 &  400 & 100 & 433 \\
\hline
Model B     &  600 &  800 & 100 & 500 \\
\hline
\end{tabular}
\caption{\label{table:model} Model spectra. }
\end{table}

It is well known that cascade decays impose restrictions on the
kinematics of the outgoing decay products.  For our gluino cascade
decay, the diquark invariant mass edge for on-shell intermediate
squarks is
\begin{equation}
\left. m_{qq} \right|_{\text{edge}} = \sqrt{ \frac{ (m_{\tilde{g}}^2 -
    m_{\tilde{q}}^2) (m_{\tilde{q}}^2 - m_{\tilde{\chi}_1^0}^2) }{
    m_{\tilde{q}}^2 }} \ ,
\label{eqn:onshelledge}
\end{equation}
assuming the quarks are massless, and where $m_{\tilde{g}}$,
$m_{\tilde{q}}$, and $m_{\tilde{\chi}}$ are the mass of the gluino,
squark, and neutralino, respectively.  For the off-shell squark case,
the edge value is simply
\begin{equation}
\left. m_{qq} \right|_{\text{edge}} = m_{\tilde{g}} - m_{\tilde{\chi}_1^0} \ .
\label{eqn:offshelledge}
\end{equation}

Of necessity, a correct assignment of the jets to the two sides of the
event will have the invariant mass of the jet-jet pair below the
kinematic edge.  Naturally, an incorrect pair assignment can produce
an invariant mass beyond the relevant kinematic edge, since the two
quarks are uncorrelated. Also, note that the diquark invariant mass
distribution through an on-shell squark possesses a characteristic
triangular shape, with its rising edge saturating the upper endpoint.
On the other hand, the off-shell squark case has the upper invariant
mass edge characteristically falling near the endpoint. Thus, the
kinematic edge is easier to identify in on-shell scenarios, and in
either case, rejection of pair assignments with invariant masses
beyond the edge serves to reduce combinatorial ambiguities.

Having defined our models, we now try and resolve combinatorial
ambiguities in these models. We used MadGraph/MadEvent
4.4.26~\cite{Alwall:2007st} to generate 100,000 events of
pair-produced gluinos at masses of 600 GeV with a 7 TeV and 14 TeV LHC
without initial state radiation.  Using BRIDGE
2.18~\cite{Meade:2007js}, we force these gluinos to decay to the
lightest neutralino through an on-shell or off-shell squark.  We do
not simulate any direct squark pair production or squark-gluino
associated production.

\section{The hemisphere method}
\label{sec:hemisphere}

The hemisphere method attempts to delineate two hemispheres in an
event, whereby all objects in a given hemisphere are ideally from the
same decay chain.  The rationale for this method is the assumption
that hard scattered parent particles are approximately back-to-back in
the lab frame.  This is not quite correct in hadron colliders because
the longitudinal parton momenta of the initial state cannot be tuned.
The hope is, however, that the  parent particles give a
large boost to their separate decay chains, and so a given cone or
hemisphere for each parent particle will generally capture all of
their daughter particles.

In practice, the hemisphere method is implemented as a two-step
process.  First, one chooses two seeds that will serve as the
starting object for each side of the event.  Second, one clusters the
remaining objects according to some figure of merit with one seed or
the other.  This figure of merit is typically taken to be $p dR$,
where $p \equiv |\Delta (\vec{p})|$ is the magnitude of the
three-momentum difference between the object and a given seed and $dR
\equiv \sqrt{ (\Delta \phi)^2 + (\Delta \eta)^2}$ is roughly the
angular separation: the object is clustered with whichever seed
minimizes $p dR$.

One also needs some method for choosing the seeds.  In our analysis,
we adopt the traditional method of choosing the highest $p_T$ object
to be the first seed.  For the second seed, following the treatment in
Sec.~13.4 of~\cite{Ball:2007zza}, we choose either (\textbf{PDR1}) the
object that maximizes $p dR$ relative to the first seed or
(\textbf{PDR2}) the object that maximizes the invariant mass of the
pair.  With these two seeds, we then calculate the $p dR$ value of the
object with respect to each seed, and we cluster the object to be with
seed 1 or seed 2 according to which seed minimizes the produced $p
dR$.

We note that since we have four quarks in each event, we have a
possibility of an assignment where one seed is by itself and three
quarks are assigned to the other side; we will refer to these as
singlet-triplet assignments.  These are all incorrect assignments,
since the proper partition of the event in the particular case we are
considering requires two quarks assigned to each side.  Though these
events do contain useful kinematic information, we choose to discard
them for simplicity.

We implement selection cuts at each stage of seed selection and object
clustering.  These cuts are summarized as follows:
\begin{my_itemize}
\item Cut 1: The $p_T$ of the initial seed quark (the
  highest $p_T$ object in the event) must be at least 200 GeV.  
\item PDR1 Cut 2: The minimum $p dR$ between seed 1 and seed 2 must be
  1800 GeV.
\item PDR2 Cut 2: The invariant mass $M$ between seed 1 and seed 2
  must be greater than the theoretically calculated diquark kinematic
  edge value (we assume this can be experimentally measured
  accurately).
\item Cut 3: We discard all singlet-triplet events, ensuring that we
  work with events that have been divided into pair-pair combinations.
\item Cut 4: We impose the restriction that the maximum seed-object
  invariant mass is the theoretical diquark kinematic edge value,
  calculated from~\eqnref{onshelledge} and~\eqnref{offshelledge}.  
\end{my_itemize}
The cut efficiency, which is the percentage of all events that survive
cuts, is summarized in~\tableref{cuts} for the above cuts.

\begin{table}[tbp]
\centering
\begin{tabular}{|l||c|c|c|c|c|c|c|c|}
\hline
 &           \multicolumn{4}{c|}{\textbf{PDR1}}
           & \multicolumn{4}{c|}{\textbf{PDR2}} \\
\hline
Model Name & Cut 1 & Cuts 1--2 & Cuts 1--3 & Cuts 1--4
           & Cut 1 & Cuts 1--2 & Cuts 1--3 & Cuts 1--4 \\
\hline
Model A - 7 TeV  & 78.8\% & 25.2\% & 12.4\% & 12.2\%
                 & 78.8\% & 51.4\% & 26.1\% & 25.7\% \\
\hline
Model A - 14 TeV & 81.7\% & 35.8\% & 18.5\% & 18.2\%
                 & 81.7\% & 58.1\% & 30.5\% & 30.1\% \\
\hline
Model B - 7 TeV  & 81.8\% & 27.1\% & 13.4\% & 13.3\%
                 & 81.8\% & 38.5\% & 19.6\% & 19.6\% \\
\hline
Model B - 14 TeV & 83.9\% & 37.5\% & 19.2\% & 18.7\%
                 & 83.9\% & 46.1\% & 24.4\% & 24.4\% \\
\hline
\end{tabular}
\caption{\label{table:cuts} Efficiencies for PDR1 and PDR2 cuts.}
\end{table}

Finally, one can place a cut on the difference of the two values of $p
dR$ calculated for each object with respect to the two seeds. The
rationale is that if the $p dR$ between an object and seed 1 is much
smaller than the corresponding $p dR$ value with respect to seed 2,
then it is more likely that the object is associated with seed 1.  By
making the minimum required difference of these two $p dR$ values
larger, the probability of an incorrect assignment can be reduced, at
the cost of losing signal events.  We therefore adopt a variable cut
that progressively obtains higher event purity at the expense of
decreasing event efficiency, which will serve as a performance measure
for later comparison.
\begin{my_itemize}
\item We change the minimum difference in $p dR$ in order for an
  object to be assigned to a given seed from 0 to 2000 GeV.  
\end{my_itemize}

\section{The $p_T$ v. $M$ method}
\label{sec:ptvm}
Our method is similar in spirit to the hemisphere method, but it is
more flexible, more model-independent, and performs better for
on-shell cascade decay chains.  We have already seen that considering
cuts in invariant mass (i.e. considering events below the diquark edge)
can reject some wrong combinations.  On the other hand, the hemisphere
method takes advantage of large $p_T$ boosts to help separate
hemispheres and reduce ambiguities.  Our goal is to combine these
ideas; simply stated, we look for correlations in high transverse
momentum pairs with low invariant mass.

\begin{figure}[tbp]
\includegraphics[scale=1.0]{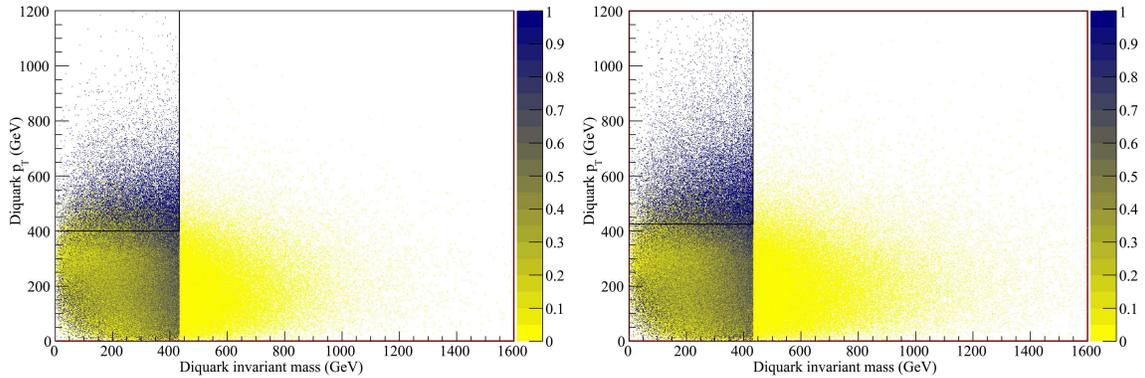}
\caption{\label{fig:modelApTvMallAB} (color online). (A) Model A - 7
  TeV; (B) Model A - 14 TeV - Diquark transverse momentum v. invariant
  mass, correct and wrong combinations.  The shading of the point
  reflects the fraction of correct combinations according to the key
  on the right.  The boxes of (A) $p_T > 400$ GeV and $M_{qq} \leq
  433$ GeV and (B) $p_T > 425$ GeV and $M_{qq} \leq 433$ GeV indicate
  regions with (A) 91.9\% event purity and 5.7\% event efficiency; (B)
  95.7\% event purity and 7.2\% event efficiency.}
\end{figure}

We begin by considering Model A at a 7 TeV $pp$ collider.  We generate
a large number of events as described above, and plot all diquark
combinations (both correct and incorrect assignments) according to
their summed $p_T$ and invariant mass $M$, in~\figref{modelApTvMallAB}.
Considering the left panel and neglecting the shading for the moment,
we notice two prominent features. The first is the kinematic edge in
the invariant mass, especially visible for high $p_T$; as we move
towards large invariant mass at high $p_T$, the number of combinations
drops off beyond a certain invariant mass. This is just a new way of
seeing that correct diquark assignments must all lie below the
knematic edge, while incorrect assignments can have much larger values
of invariant mass, producing an excess of events at lower invariant
mass.  Correspondingly, the excess of diquark combinations with
invariant mass larger than the edge value must all be wrong
combinations.

We also recognize a second feature; as we increase $p_T$, there is
again an excess in the diquark combinations below the kinematic edge.
This is particularly noticeable if we only look at the events which
have invariant masses larger than the kinematic edge (which are all
incorrect combinations). These events tail off quickly towards higher
$p_T$, while events below the kinematic edge (a mixture of correct and
incorrect combinations) extend to higher $p_T$.  A natural guess
(based on the success of the hemisphere method) is that the
combinations at large $p_T$ are dominated by correct assignments.
Roughly, we expect correct combinations to carry higher $p_T$ than
wrong combinations because correct combinations share the same
transverse momentum of the parent gluino in the lab frame, while wrong
combinations should generically have canceling parent gluino
transverse momenta.

\begin{figure}[tbp]
\includegraphics[scale=1.0]{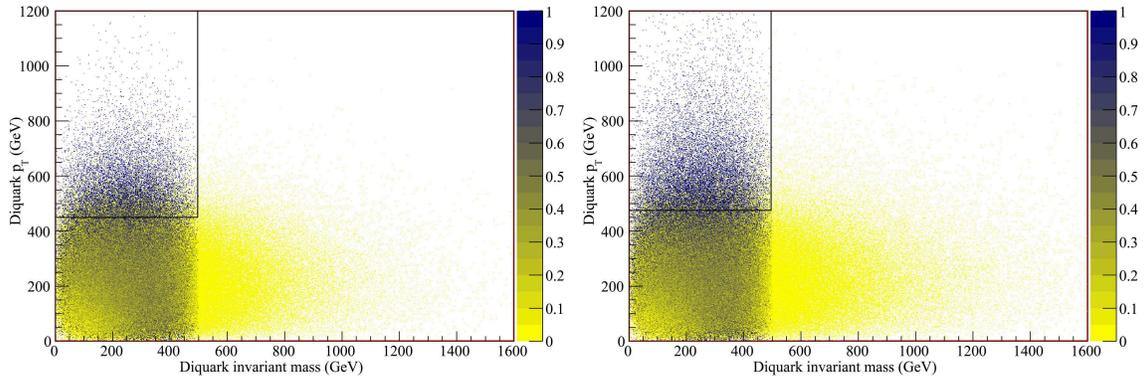}
\caption{\label{fig:modelBpTvMallAB} (color online). (A) Model B -
    7 TeV; (B) Model B - 14 TeV - Diquark transverse momentum
    v. invariant mass, correct and wrong combinations.  Same shading
  scheme as~\figref{modelApTvMallAB}.  The boxes of (A) $p_T > 450$ GeV
  and $M_{qq} \leq 500$ GeV and (B) $p_T > 475$ GeV and $M_{qq} \leq
  500$ GeV indicate regions with (A) 91.7\% event purity and 3.1\%
  event efficiency; (B) 94.4\% event purity and 4.2\% event
  efficiency.}
\end{figure}

We have used our knowledge of the parton-level information to
shade~\figref{modelApTvMallAB} according to the percentage of diquark
pairs at each $p_T$ v. $M$ point that are correct.  We see explicitly
that the correct and incorrect combinations occupy distinct regions of
this plot; the correct combinations characteristically have larger
$p_T$ and lower invariant masses.  Moreover, as we increase to a 14
TeV LHC, we expect the population of high $p_T$ correct combinations
to increase, because of the higher boost of the parent particles, as
depicted in the right panel of~\figref{modelApTvMallAB}.  Similar
results are also evident for Model B, as seen
in~\figref{modelBpTvMallAB}.

This strongly suggests that a cut on combinations with invariant mass
below the kinematic edge and high $p_T$ can guarantee an event sample
dominated by correct combinations.  We now present a practical method
for performing this cut.

\section{The $p_T$ v. $M$ procedure}
\label{sec:ptvmprocedure}

Given an event with four quarks, we implement our method
according to the following procedure:
\begin{my_enumerate}
\item For each event, plot the $p_T$ and $M$ of each diquark pair (all
  six pairings).
\item Isolate the diquark invariant mass edge $M_0$.  This is usually
  immediately visible as a cutoff at larger invariant masses.
\item Now only consider the pairs with invariant mass larger
  than the kinematic edge.  Choose a $p_T$ threshold value $P_0$ such
  that fewer than 5\% of these pairings have $p_T$ larger than this
  threshold value.  For example, in Model A at a 7 TeV LHC, this
  corresponds to a threshold value of about 400 GeV.
\item For each event, we require that there be a division of the 4
  quarks into two pairs such that the $p_T$ of each pair is larger
  than $P_0$, and the invariant mass of each pair is below $M_0$. If
  there is no such division possible or if more than one division is
  possible, we discard the event. If exactly one division passing the
  requirement is possible, the event is retained and the assignment
  which passes the cut is treated as the proper assignment of the
  quarks to the two sides.
\end{my_enumerate}

We note that neither the $p_T$ cut nor the invariant mass cut we have
chosen have been optimized.  Choosing a higher $P_0$ would generally
ensure a purer event sample with fewer incorrect combinations;
however, a higher $P_0$ also reduces signal statistics. Finding an
optimum balance between event purity and signal statistics will be a
model-dependent, search channel-dependent question.  Moreover, this
procedure is readily extended to more complicated event signatures,
such as multi-jet and multi-lepton events, and can also be used as a
figure of merit to cluster entire sides all at once.

\section{Comparison of Methods}
\label{sec:comparison}
In this section, we present a comparison between the hemisphere method
and our new $p_T$ v. $M$ method.  We will use event purity against
event efficiency as a comparison measure.  We will define event purity
as the percentage of events where the quarks of the event are paired
correctly.  Event efficiency is defined as the percentage of total
events that pass the imposed cuts.

We can benchmark the performance of the hemisphere method and the
$p_T$ v. $M$ method by adjusting cuts.  For the hemisphere method,
increasing the minimum difference in $p dR$ in order for an object to
be assigned to a seed makes it more likely that a given object and
seed are from the same decay chain.  For the $p_T$ v. $M$ method, we
can increase the success of isolating the correct pair-pair
combination by increasing the $p_T$ cut on each pair of a pair-pair
combination.

\begin{figure}[tbp]
\includegraphics[scale=1.0]{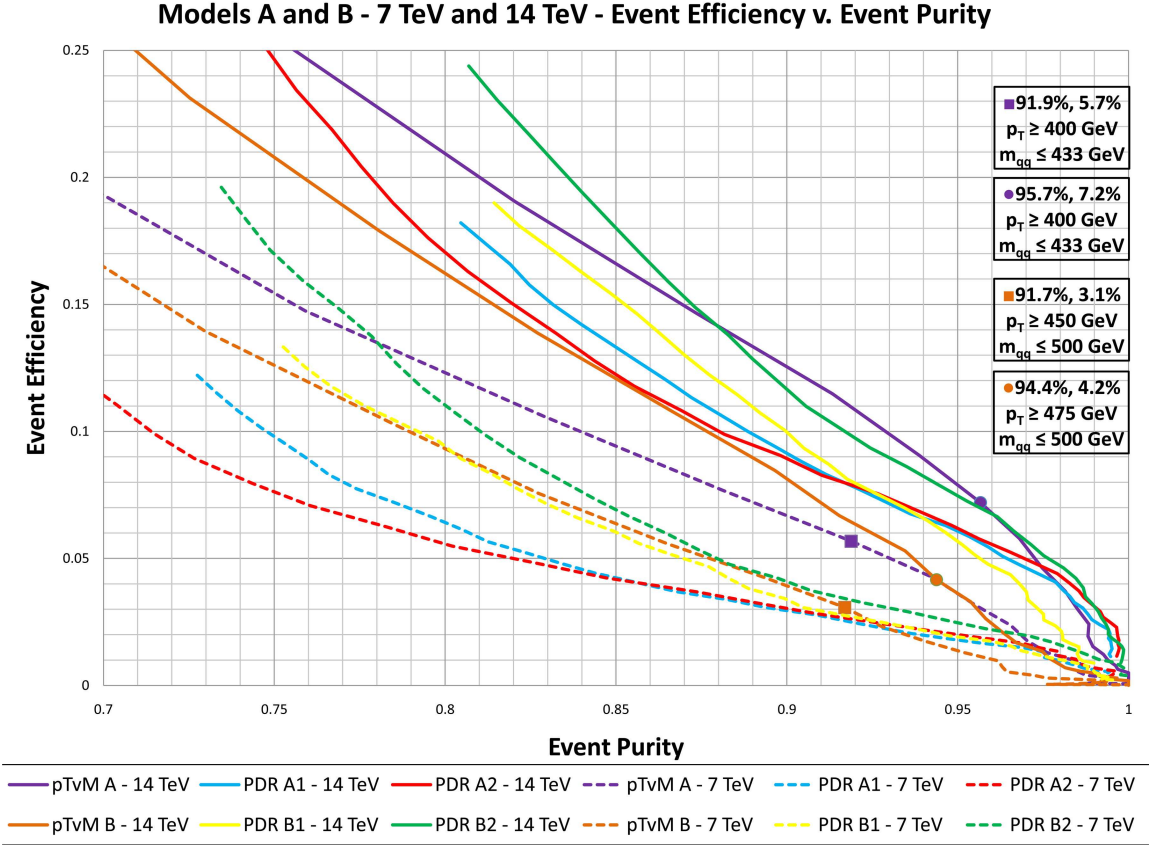}
\caption{\label{fig:modelab_evt} (color online). Models A and B - Event
  purity v. efficiency, as defined in the text.  The marked points of
  (91.9\%, 5.7\%) and (95.7\%, 7.2\%) for Model A and (91.7\%, 3.1\%)
  and (94.4\%, 4.2\%) for Model B correspond to the outlined boxes
  in~\figref{modelApTvMallAB} and~\figref{modelBpTvMallAB}, respectively.}
\end{figure}

In~\figref{modelab_evt}, we show the event purities as a function of
event efficiencies for Model A with a 7 TeV and 14 TeV LHC.  For this
on-shell mass spectrum, we can readily see that the $p_T$ v. $M$
method delivers far better event efficiency for a given event purity
than either PDR1 or PDR2.  For 85\% purity, the Model A - 7 TeV event
sample from $p_T$ v. $M$ is approximately twice as large as either
hemisphere method sample.  While the difference in performance is
moderately reduced for a 14 TeV LHC, the superiority in performance is
still impressive.  At 90\% purity, for example, the $p_T$ v. $M$ event
sample is about 33\% more.  From our earlier procedure given
in~\secref{ptvmprocedure}, where the survival rate of wrong diquark
pairings in the region to the right of the invariant mass edge was
about 5\%, we obtain the two marked points of (91.9\%, 5.7\%) for
Model A - 7 TeV and (95.7\%, 7.2\%) for Model A - 14 TeV.

For the off-shell scenario, also shown in~\figref{modelab_evt}, the
performance for all three methods is approximately the same at a 7 TeV
LHC.  A 14 TeV LHC, however, shows the PDR2 method performing best,
with a 50\% gain in event efficiency over the $p_T$ v. $M$ method for
85\% event purity.  The difference is greater for higher purities,
becoming approximately 100\% for 95\% purity.  Using the procedure
of~\secref{ptvmprocedure} with a survival rate of about 5\%, we find
the points (91.7\%, 3.1\%) and (94.4\%, 4.2\%) for Model B - 7 TeV and
14 TeV, respectively.  While good purity can be obtained for off-shell
models using the $p_T$ v. $M$ method, the hemisphere method is better.
We can hypothesize that the hemisphere methods work better in the
off-shell case because the directionality of the diquark pairs is
roughly uniform.  Because the hemispheres created using $p dR$ can lie
in an arbitrary orientation with respect to the beam axis, they are
more flexible in capturing the correct diquark pairings than the $p_T$
v. $M$ method, which necessarily requires the two quarks to have lots
of momenta in the plane transverse to the beam axis.  When the
intermediate squark is on-shell, however, the squark typically gives a
large boost to the second emitted quark, which helps result in a high
$p_T$ diquark pair.  From the plots, we thus observe that the
traditional $p dR$ method of assigning particles to decay chains seems
well-suited for off-shell decays, but is significantly outperformed by
our new $p_T$ v. $M$ method for on-shell cascade decays.

\section{Conclusion}
\label{sec:conclusion}

We have compared the performance of the hemisphere method with the
$p_T$ v. $M$ method using the models presented in~\tableref{model}.
We have shown the $p_T$ v. $M$ selection criteria
of~\secref{ptvmprocedure} delivers $\mathcal{O}(25\%)$ to
$\mathcal{O}(100\%)$ more event statistics for a given purity between
$85\%$ to $95\%$ than the traditional $p dR$ methods for an on-shell
decay chain.  This is a substantial increase in the number of viable
signal events and would be a useful method for any study that suffers
from decay chain assignment combinatorial ambiguities.  The hemisphere
method, on the other hand, seems well-suited for off-shell decay
chains.  A realistic collider study should benefit from adopting both
methods in parallel, since it is not generally known beforehand
whether the observed decay products are produced on-shell or
off-shell.  One possibility for improving the off-shell performance of
$p_T$ v. $M$ is to salvage the events with more than one combination
(not ``pure'') by retaining only the pair-pair combination that has
the largest scalar total $p_T$.  This is left for future work.

We remark that although this study was performed using simplified
models without detector effects, ISR/FSR, or background simulation, we
believe the major features of our results are retained when going to
full models.  As an important validation, our hemisphere method
performance results are qualitatively similar to those presented in
Sec.~13.4 of~\cite{Ball:2007zza}, which used a CMS detector
simulation.  At high LHC energies, however, the effects of ISR is
clearly not negligible, especially when the ISR jet gives a large
$p_T$ boost to the entire hard process.  In such scenarios, we would
expect the additional ISR jet to adversely degrade both efficiency and
purity of retained signal events.  We expect the effect to be similar
to both the hemisphere method and the $p_T$ v. $M$ method, though,
since the inclusion of a large $p_T$ ISR jet makes the remaining jets
more isotropic.  We are currently studying possible implementations of
the $p_T$ v. $M$ method to tag events with large $p_T$ ISR jets and
potentially tag ISR jets.

One drawback of our method is the assumed observation of the dijet
invariant mass edge.  While we do not prescribe an exact method to
extract this edge largely because of jet energy scale uncertainties,
our method highlights the fact that wrong combinations (which can
pollute and exceed the edge value) can be removed with moderate dijet
$p_T$ cuts, perhaps allowing a more accurate extrapolation of the edge
value if one understands the expected shape of the resulting invariant
mass distribution.  Moreover, mass measurements initially will focus
on dilepton invariant mass edges, which are relatively straightforward
to observe given the good understanding of lepton kinematics in
detectors.  We envision that once dilepton edges are measured, the
$p_T$ v. $M$ method could be used to isolate any available
dilepton+jet and dilepton+dijet edges; given this information, the
edge in the dijet cross channel could be extracted and further
kinematic studies in multi--jet final states would proceed given the
relatively pure signal samples determined from the $p_T$ v. $M$
method.

A separate issue is that our on-shell Model A spectrum, which was
chosen to mimic the SPS1A benchmark point, gives distinct, highly
energetic jets.  If the jets were less energetic, reflecting a
compressed SUSY spectrum, (for example, $\tilde{g} = 600$ GeV,
$\tilde{q} = 500$ GeV, and $\tilde{\chi}_1^0 = 400$ GeV) we would find
that the $p_T$ v. $M$ method performs approximately as well as the
hemisphere method at a 7 TeV LHC and slightly worse (about 15\%--25\%
less efficient) for a 14 TeV LHC, but the overall efficiency is
increased by a factor of 2 for the $p_T$ v. $M$ method.  For
compressed spectra, we thus see the relative performance of the $p_T$
v. $M$ method matches that presented for the off-shell Model B
spectrum.

We expect the $p_T$ v. $M$ method to be useful for other signatures
besides the one presented in this study, since we are taking advantage
of kinematic features found in all cascade decay chains.  It is
straightforward to consider combinatorial ambiguities arising from
multi--lepton or leptons+jets final states.  Additionally, we could
extend our method to resolve ambiguities when the final state includes
heavy, reconstructed $W$ or $Z$ gauge bosons with minor modifications
to the expected invariant mass spectrum.  One main difficulty with
such a generalization, however, is multiple branching modes occur with
similar kinematics, such as a gluino decay via multiple on-shell
squark modes where the squarks are largely degenerate.  Here, the
signal-only combinatorial ambiguity among light jets is shared among
multiple decay modes, and because the observed kinematics are similar,
it will be difficult to apply kinematic cuts to distinguish distinct
decay chains.  We are currently in the process of applying this method
to analyze new physics models passing through a realistic detector
simulation.  In particular, a study of precision determination of
invariant mass edges including dilepton and dijet edges at the SPS1A
benchmark point is underway and seeks to demonstrate the usefulness of
this method in a collider environment.

A further direction of research is to improve the $p_T$ v. $M$ method,
for example by performing a shape analysis of the correct vs. wrong
combinations in the $p_T$ v. $M$ plane in order to optimize the $p_T$
v. $M$ cut.  It may also be possible to implement the $p_T$ v. $M$
method on an event-by-event basis, for example, by preferentially
choosing pair-pair combinations at the event level according to the
maximum $p_T / M$ or another similar figure of merit.  Such a measure
may be superior to the $p dR$ method for off-shell decay scenarios.
Finally, we would also want to study how our method can be applied to
improve the determination of the overall mass scale and the LSP/LKP
mass at colliders.  We hope to return to these issues in future work.


\section*{Acknowledgements}
\label{sec:acknowledge}

FY is supported by an LHC Theory Initiative Graduate Fellowship, NSF
Grant No. PHY-0705682.  AR and FY are supported by NSF Grant
No. PHY-0653656 and PHY-0970173.


\end{document}